\documentstyle[11pt,aas2pp4,psfig]{article}

\received{}
\accepted{}
\accepted{}
\cpright{AAS}
\ccc{}

\slugcomment{}

\lefthead{Binette et~al.}
\righthead{Fermi acceleration of Lyman $\alpha$}

\newcommand{\cms}{\hbox{\,${\rm cm^{-2}}$}}

\newcommand{\cmc}{\hbox{\,${\rm cm^{-3}}$}}
\newcommand{\kms}{\hbox{\,${\rm km\,s^{-1}}$}}

\newcommand{\SS}{{\S}\,}
\newcommand{\etal}{\hbox{et~al.}}
\newcommand{\eg}{\hbox{e.g.}}
\newcommand{\cf}{\hbox{cf.}}

\newcommand{\degree}{\hbox{$^{\sc o}$}}
\newcommand{\vsep}{\hbox{$v_{sep}$}}
\newcommand{\vs}{\hbox{$V^{s}_{7}$}}
\newcommand{\vpk}{\hbox{$V_7^{peak}$}}
\newcommand{\vtur}{\hbox{$v_{turb}$}}
\newcommand{\nsou}{\hbox{$N_{21}^{sou}$}}
\newcommand{\nmir}{\hbox{$N_{21}^{mir}$}}

\newcommand{\nprec}{\hbox{$N_{21}^{prec}$}}
\newcommand{\tsou}{\hbox{$T_{2}^{sou}$}}
\newcommand{\tmir}{\hbox{$T_{2}^{mir}$}}

\newcommand{\mud}{\hbox{$\mu_{\sc D}$}}

\newcommand{\civ}{\hbox{C\,{\sc iv}}}

\newcommand{\lya}{\hbox{Ly$\alpha$}}

\newcommand{\hi}{\hbox{H\,{\sc i}}}
\newcommand{\hii}{\hbox{H\,{\sc ii}}}

\begin{document}

\title{Evidence of Fermi
acceleration of Lyman $\alpha$ \\ in the Radio Galaxy
1243+036\footnote{Based on NTT observations archived at the European
Southern Observatory: http://arch-http.hq.eso.org/ } }

\author{Luc Binette\altaffilmark{2}, Benoit Joguet\altaffilmark{3} 
and John C. L. Wang\altaffilmark{4} } 
\affil{binette@astroscu.unam.mx, bjoguet@eso.org, jcwang@astro.umd.edu}

\altaffiltext{2}{Instituto de Astronom\'\i a, UNAM, Ap. 70264, 04510
M\'exico, DF, M\'exico}

\altaffiltext{3}{European Southern Observatory, Casilla 19001, 
Santiago 19, Chile}

\altaffiltext{4}{Astronomy Department, University of Maryland,
College Park, MD 20742-2421, USA}

\begin{abstract}

The high redshift radiogalaxy 1243+036 ($z=3.6$) presents an
asymmetric \lya\ profile of FWHM 1550 \kms\ as measured by van~Ojik
\etal\ We propose that the blue asymmetry in the \lya\ profile is not
due to narrow absorption dips but consists of narrow emission peaks.
We interpret the blueshifted peaks near $-1130, -850$ and $-550 \kms$
(relative to the peak of full profile) as being the result of Fermi
acceleration of \lya\ produced by jet-induced star formation in the
wake of a 300~\kms\ shock. This shock would be caused by the
deflection of the radio-jet at the observed position of the radio bend
which also coincides spatially with the excess \lya\ emission reported
by van~Ojik \etal\
\end{abstract}

  



\section{Introduction} \label{intro}

Lyman $\alpha$ is the strongest line observed in very high redshift
radio galaxies (HZRG). Although the brightness of \lya\ peaks near or
at the nuclear position, most of the emission is spatially resolved
with the fainter emission extending up to radii 40--130\,kpc. The
\lya\ profile in HZRG is characterized by a FWHM in the range
700--1600\kms\ (van~Ojik 1995 and references therein).  The
intermediate resolution study of \lya\ profiles carried out by
van~Ojik \etal\ (1997) has revealed the presence of troughs which are
well explained by \hi\ gas absorption present in the environment of
the parent radio galaxy. In their sample, however, the HZRG 1243+036
($z=3.6$) may call for a different interpretation, namely that the
broad profile presents true narrow emission features in between the
narrow `dips' which van~Ojik \etal\ (1996: vO96) have interpreted as
absorption features. The repeated scattering of the resonance \lya\
line across a shock discontinuity was shown by Neufeld \& McKee (1988:
NM88) to result in a systematic blueshift of \lya.  In this paper, we
develop in more detail the Fermi acceleration model and propose that
the narrow features observed by vO96 correspond to a small number of
across-shock scatterings.

\section{The model} \label{model}

The calculations of the \lya\ profile is carried out in two steps. First
we compute the emergent profile from the ionized gas, second we
consider the effect of Fermi acceleration across a shock discontinuity
which is taking place between the observer and the \lya\ emission gas. 

\subsection{Double-horned source profile} \label{sour}
A resonant photon escaping from a static slab will diffuse in
frequency and space until it can escape in the wings of the opacity
profile. This leads to a symmetric double-horned profile with a
separation, \vsep, between the two peaks of:

\begin{equation}
v_{sep} \approx 390 \left(N^{sou}_{21} \sqrt{T_2^{sou}}\right)^{1/3} ~{\rm km~s^{-1}}\;,  \label{eqvsep}
\end{equation}

\noindent where $10^{21}\nsou\ \cms\ $ is the neutral hydrogen column
density surrounding the emitting ionized gas and $10^2 \; \tsou $~K
its temperature.  We properly take into account the skewness at large
\nsou\ of the two emission peaks (\cf\ Bonilha \etal\ 1979) by using
the numerical prescription described in Appendix~B of Binette \etal\
(1993). We can either assume the ionized gas to be homogeneous and
isothermal (structure-less) or we can integrate the \lya\ profile
escaping a cooling shock slab or a stratified photoionized slab. In
either case, we assume a plane-parallel geometry for the transfer
which allows two possible perspectives for an external observer: front
or back. The front perspective corresponds to the precursor side
(shock wave case) or the UV source side (photoionization case). The
fraction of \lya\ escaping towards the front or the back can be quite
different, especially for an ionization bounded photoionized slab or
for a shock wave with a trailing thick shell. The \hi\ gas beyond the
Str\"omgren boundary or within the trailing shell acts like mirrors
which favor \lya\ photons to escape overwhelmingly towards the
front. This effect as well as \lya\ destruction due to internal dust
are fully taken into account using the formalism developed in
Appendix~A of Binette \etal\ (1993). We label \nmir\ the column
density of the trailing \hi\ shell.
 
\subsection{Fermi acceleration of resonant photons} \label{fer}
As for the effect of Fermi acceleration on the \lya\ profile, unless
stated otherwise we will consider the front perspective whereby the
shock discontinuity is taking place between the observer and the
region producing the line. This is different from the (back)
perspective\footnote{Our front perspective prevents the components
$(a)$ and $(b)$ discussed in NM88 from being visible.} in NM88 in
which the shock was on the far-side of the emergent \lya\ (see their
Fig.~1). As in NM88, we treat the slabs of scattering material on
either side of the shock front \nprec\ and \nmir\ as two converging,
partially reflecting mirrors. Each passage of a resonant photon across
the shock front corresponds to a reflection and, for an isotropic
radiation field, the mean blueshift after $n$ passages across the
discontinuity is $nV_s/2$, where $V_s$ is the shock wave velocity.  If
the transmissivity of the mirrors on the $k$th reflection is $T_k$ and
the reflectivity is $1-T_k$, the fraction of resonant photons escaping
at exactly $n$ shock crossings is

\begin{equation}
f(n) = T_n \prod_1^{n-1}{\bigl(1-T_k\bigr)} \approx T_n {\rm exp}\Bigl(-\int_0^n{T_k \;
dk} \Bigr)  \;.  \label{eqsurv}
\end{equation}
\noindent The right term is a valid approximation only in the limit of
large $n$ and $T_k\ll 1$. Unlike NM88, we have used instead the middle
term in which each crossing of the discontinuity is calculated
following the algorithm described in Appendix~A. The main advantage is
that it allows us to cover the case of a small number of
reflections. Furthermore, it retains higher resolution of the spectral
features appearing on the transmitted profile and it allows a more
accurate treatment of the effect of dust absorption (see
Appendix~A). We define \mud\ as the dust-to-gas ratio relative to the
local interstellar value.

The profile of non-resonant lines of species $i$ of atomic mass $m_i$ are
assumed gaussian with a FWHM given by $1.665\left( 2kT/m_i + v_{turb}^2
\right) ^{1/2}$ where \vtur\ is the local `most probable' turbulent
velocity. This summarizes the implementation of profile determination
in the multipurpose code {\sc mappings~ic} (\cf\ Ferruit
\etal\ 1997).

\subsection{Profiles produced by Fermi acceleration}\label{explo}

To illustrate the type of profiles obtained, we have made calculations
under a wide range of input parameters. In these calculations, the
shock velocity, $100 \; \vs\ \kms\ $, is constant at $\vs =
3.0$\,. Increasing \vs\ does not necessarily change the shape of
profile if the opacities are increased accordingly (although it does
blueshift the profile further). For definiteness, we have set $\nprec
= \nmir $ in all our calculations\footnote{In the event of a large
imbalance, clearly \lya\ would escape in the direction of the thinnest
of the two mirrors.}. We adopted a temperature of 100\,K for the gas
responsible for the scattering. The choice of temperature impacts only
on \vsep\ [see equation~(1)] and therefore only the product
$N^{sou}_{21} \sqrt{T_2^{sou}}$ matters to the calculations. At constant
temperature, the two essential free parameters are therefore \nmir\
($=\nprec$) and \nsou\ [the column depth immediately surrounding the
ionized gas, see equation~(1)]. The different profiles resulting from
Fermi acceleration are presented in Fig.~1 and the corresponding input
parameters in Table~1.  It is seen from Fig.~1 that the higher \nmir,
the higher the blueshift of the emergent Fermi profile (see Appendix~B
for a comparison with NM88). \vsep\ is an important factor governing
the appearance of the profile when it is $\ga V_s$. The effect of dust
on Fermi acceleration is not as severe as in NM88. The main reason is
that we considered only the effects of dust within the two mirrors and
its implication on the Fermi profile {\it per se} while disregarding
the presence of dust within \nsou, the gas immediately surrounding the
source. In any event, little dust absorption would take place when
\nsou\ is small, a possibility strongly favored in the discussion
below. Also we expect the extremely hot postshock gas surrounding the
embedded condensations to be devoid of dust. In our calculations of
the dust effects on the Fermi profile, we have taken into account the
mean geometrical depth traversed at each shock crossing by the
resonant photon. Fig.~1$j$ illustrates the case with $\mud=0.2$ (dust
content of mirrors only) which should be compared with the dustfree
case of Fig.~1$i$.

\section{Analysis of the 1243+036 data} \label{redu}

We have reduced the ESO archived observations of PKS~B1243+036 taken
at the NTT with EMMI by vO96 on 14 April 1994. The 4 one hour duration
long-slit spectra with PA=152\degree\ (aligned with the radio axis)
were registered on a $2048\times 2048$ Tektronix CCD. The $2\times 2$
binning gave a spatial scale of $0.54\arcsec
{\rm pix}^{-1}$. Grating~6 was used with a wide slit of
2.5\arcsec. Our analysis of the arc spectra indicates a resolution
(FWHM) of $\simeq 2.6$\AA.  


After performing a flat-field division, sky subtraction, flux
calibration and correction for atmospheric extinction using the IRAF
reduction package, we extracted a one-dimensional spectrum using a
restricted window of 7\arcsec, centered on the bending point of the
radio source (\cf\ Fig.~4 and 14 of vO96). The resulting spectrum is
plotted as solid line in the upper frame of Fig.~2. It differs from
that of vO96 as a result of considering a window which is both smaller and
centered around the radio bend.

Several small peaks (present in the individual exposures) appear on
the \lya\ profile at 5536, 5541 and 5547\AA.  In order to isolate
these features, we smoothed the spectrum and subtracted the result
from the original spectrum.  The result is plotted as a dotted line in
the upper frame of Fig.~2.  Obviously, the negative part of this curve
is unphysical whether the features are emission peaks or absorption
dips. To remedy this, we supposed that the underlying \lya\ profile is
intrinsically symmetric since almost all the \lya\ profiles measured
by van~Ojik \etal\ (1997) turned out to be symmetric {\it after}
correction of the absorption troughs.  Taking the red side of the {\it
smoothed} profile as our reference, we folded it around the peak
position ($\lambda_{Max}=5557.2$\AA) over the blue side and subtracted
it from the original profile.  The resulting spectrum is shown in the
lower frame of Fig.~2 (dotted line).

\section{Fermi acceleration in 1243+036\,?} \label{fit}

Are the high frequency features (dotted line in Fig.~2b) absorption
dips or emission peaks? The absorption interpretation is ambiguous
because the V-shaped dips are unresolved in wavelength, i.e., their
widths are less than the width of the flat-topped instrumental
profile. This can be seen in Fig.~2b where the instrumental profile is
shown in the inset box.  On the other hand, if interpreted as emission
features, these features appear to fit well the instrumental profile,
as is evident from Fig.~2b.  Furthermore, the peaks are equally spaced
in wavelength and considerably blue shifted relative to systemic
velocity, all in accordance with Fermi acceleration.  Clearly, higher
spatial and spectral resolution spectra would help settle the
question.  The existence of a shock is not in doubt, however, since
1243+036 clearly shows the presence of a radio bend at a distance of
2\arcsec\ from the nucleus (13.6\,kpc for $H_0=50\; {\rm
km\;s^{-1}Mpc^{-1}}$ and $q_0=0.5$), that is at the same spatial
position where vO96 reports genuine enhanced \lya\ emission
(underlying their absorption dips).

Let us thereafter assume that the narrow emission peaks are
independent from the underlying symmetric \lya\ profile and that they
are emitted near or at the radio bend.  Between successive peaks, two
reflections must necessarily take place. Therefore the velocity of the
shock in our model is fixed by the distance between peaks which gives
$\vs \simeq 3.0$\,.  If the emission peaks corresponded to gas
photoionized by the nucleus, we could {\it not} explain these by Fermi
acceleration since one of the necessary \hi\ mirrors (at the back)
could not exist between the ionized gas and the source. Our Fermi
acceleration model therefore requires {\it in situ} photoionization,
either by the shock itself (a) or by an independent local source (b).

In case~a, the post-shock temperature is of order $1.5 \times
10^{6}\;$K and in order to reproduce the integrated luminosity of the
emission features ($2.4 \times 10^{43} {\rm erg} \; {\rm s^{-1}}$), 
one requires the uncomfortably high pre-shock density of
$n_0=3\times10^6$\cmc\ for a shock front area as large as $10^4\; {\rm
pc}^{2}$ (we included the ionizing photons generated within the shocked
gas). Furthermore, the abundances of the heavy elements must be
extremely low since \civ\ is absent from the spectra (vO96). When we
compute the \lya\ profile from a shock, it turns out significantly wider
(after convolution) than the observed narrow peaks.

In case~b, \lya\ emission originates from starbursts triggered by the
passing of the shock wave which is the result of the deflection of the
radio jet. This possibility goes along the suggestion of Rees (1989)
that enhanced star formation can be triggered by the interaction of
the radio jet with the environment of HZRG. By inspection of many
profiles such as in Fig.~1, two possibilities arise.  One is that the
\hii\ gas is embedded in thick neutral gas such that \vsep\ is equal
to $V_s$. In this particular case, the convolved emission peaks are
sufficiently narrow. Such a model is shown in Fig.~1$l$ and it gives,
after convolution, an excellent fit.  A less {\it ad hoc} situation
which does not rely on having $\vsep=V_s$ is to have the \hii\ regions
{\it not} fully embedded but rather in their champagne phase
(Tenorio-Tagle 1979). This geometry which facilitates considerably the
escape of \lya\ into the surrounding shocked gas results in a small
\vsep. Such a nebular model was computed using a stellar atmosphere
model of 40\,000\,K. The Fermi profile obtained gives an encouraging
fit as shown by the solid line in Fig.~2b. It was computed using
$\vs=3.0$ and $\nmir=1.3$ and has a very similar profile to the
simpler model of Fig.~1$k$ when the latter is convolved. A point in
favor of the Fermi interpretation is that the peaks appear at $-4, -3,
-2 \; V_s$ relative to the adopted value of line center at 1.3\AA\
($70 \kms \simeq \frac{1}{4} V_s$) to the red of $\lambda_{MAX}$, while
absorption dips ought to be distributed randomly in velocity space.

It is interesting to note that the shock crossing time across a
projected distance of 0.2\arcsec\ --an upper limit to the projected
radio hot spot size-- is $\sim 4 \times 10^{6}$ years.  Incidentally,
a preshock gas density of 0.16\cmc\ [eight times higher than in Rees
1989] with a column of $\nmir=1.3$ possess a geometrical depth of
0.2\arcsec. After being shocked, such gas would cool on a timescale of
$\sim 0.8 \times 10^{6}$ years from a postshock temperature of
$1.5\times 10^6\;$K. If the flow became very chaotic after the gas has
cooled, this might preserve a small value for \vsep\ even if the \hii\
regions were to turn on at a later stage further downstream.

Interestingly, the only other candidate HZRG for which Fermi
acceleration was proposed, 3C326.1 (NM88), is similarly devoid of
\civ\ emission, a fact difficult to reconcile with either nuclear
photoionization by a powerlaw or with high velocity (photoionizing)
shocks. Jet-triggered star formation as we argued for above
would solve this conundrum since \civ\ is then expected to be very weak
even for solar metallicity.

\acknowledgments

We wish to thank Huub R\"ottgering for directing us to the archived
NTT data of 1243+036 which were reduced a second time by one of us
(BJ). We also thank Irapuan Rodrigues de Oliveira for an initial quick
survey on the type of profiles produced by Fermi acceleration.
One of us (JCLW) acknowledges support by NASA Astrophysics Theory Program
grant NAG5-3836 and by a Graduate Research Board award from the University
of Maryland.

\appendix

\section{Numerical algorithm to calculate Fermi profiles}

We discretize all profiles in vectors whose indices are proportional
to the velocity shift ($v_i = i \; \delta v $) from line center
$\lambda_0$. $\delta v$ is a proportionality constant sufficiently
small to allow fair sampling of the source profile. The source line
profile emergent at the shock discontinuity is defined as
$\varphi(v_i)$ (\cf\ \SS{\ref{sour}). It is centered and symmetric
around $v_i=0$. In this scheme, the observer is either on the side of
the precursor (`front' perspective): $p=+1$, or downstream beyond the
trailing \hi\ shell (`back' perspective): $p=-1$. After $j+1$ passage
across the discontinuity of the photons coming from position $v_i$ in
the source profile, their {\it contribution} to the observed profile
$\varphi^p(v_k)$ for an observer whose perspective is $p=(-1)^j$ is
given by:

\begin{equation}
 \varphi(v_i) T(v_j) \prod_{n=0}^{j-1}{\Bigl(1-T(v_n)\Bigr)} \;,  \label{eqmul}
\end{equation}
\noindent where $k$ is the nearest integer to
$v_j/\delta v$, and with $v_j$ and $v_n$ given by:

\begin{equation}
v_j = v_i - j V_s/2 - (-1)^j V_s \;,  \label{eqvk}
\end{equation}
\begin{equation}
v_n = v_i - n V_s/2 - (1+(-1)^j) V_s/2 \;.  \label{eqvn}
\end{equation}
\noindent The profiles $\varphi^{+1}$ and $\varphi^{-1}$ are obtained
by adding contributions from all $j$ until $1-T(v_j)$ becomes
negligible (note that $j$ starts at zero). The transmissivities are
approximated as $T(v_n) = [ 1+ 0.75 \tau(v_n) ]^{-1}$.  Assuming a
Holtsmark function (Lang 1974), $\tau(v_n)$ is the line opacity away
from line center (by the amount $\lambda_0 v_n/c$) due to the opacity
of either the precursor ($n$ even) or of the shell ($n$ odd).  The
line center opacities $\tau_0^p$ are derived from \nprec\ (if $p=+1$)
and \nmir\ (if $p=-1$). The attenuation of the transmissivity due to
dust absorption has been taken into account by determining the column
density of dust crossed by the resonant photon at each passage across
the discontinuity (assuming that reflection takes place at a
geometrical depth such that $\tau(v_n)\approx 1$).  As in NM88, we
neglect the diffusion in frequency that results from scattering within
the static material on either side of the shock front.

\section{Comparison with NM88}

The position of the peak intensity within the Fermi accelerated
profile, $V^{peak}$, shifts towards the blue as \nmir\ and/or \vs\ are
progressively increased.  Our numerical calculations 
lead to the following relation valid only when $N^{mir}_{21} \gg 0.002
(V_7^s)^2$:

\begin{equation}
V^{peak}_7 \simeq 5.82 \; (1+\delta) \left(V_7^s N^{mir}_{21}\right)
^{\frac{1}{3}}\;, 
\label{eqvpk}
\end{equation}
\noindent with $\delta$ defined as:
\begin{equation}
\delta = 0.16 \; (N^{mir}_{21})^{-\frac{1}{3}} \; \sqrt{V_7^s} \;,
\end{equation}
\noindent where $100 \; \vpk\ \kms$ is the peak intensity
velocity. This expression assumes that $\nprec = \nmir$.  When $\nmir
\ga 10^{4}$ (that is, for a negligible $\delta$), the above expression
for \vpk\ becomes identical to that of NM88. A more general expression is
to appear in Joguet \& Binette (1998).



\begin{figure}
\figurenum{1}

{\psfig{figure=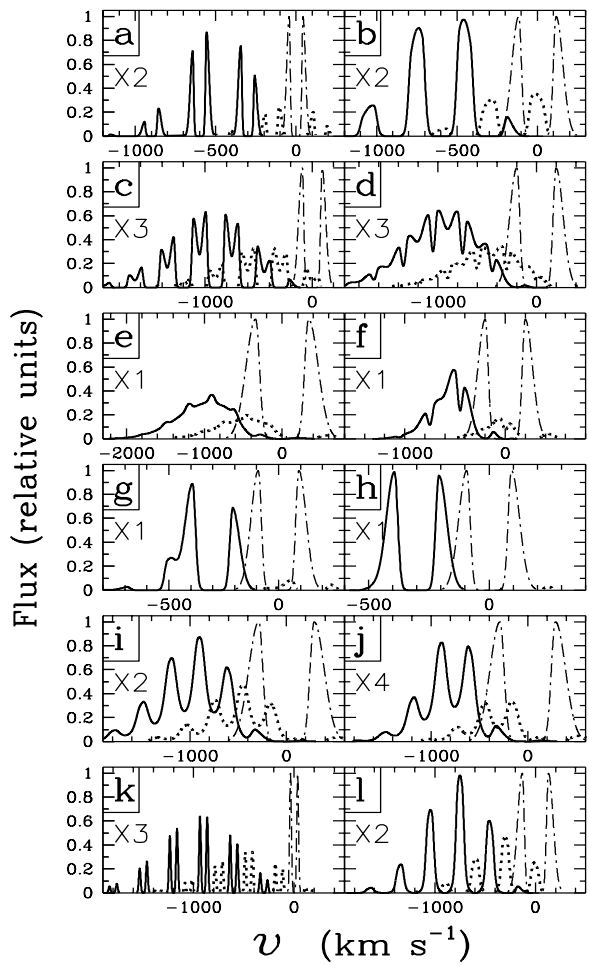,height=14cm,width=15.3cm}}
\figcaption{The dash-dotted line represents different double-horned
source profiles (\cf\ \SS\ref{sour}) as a function of velocity from
line center ($v=c \; (\lambda-\lambda_0)/{\lambda_{0}} $). The
profiles resulting from Fermi acceleration were calculated using the
parameters given in Table~1. After multiplication by an appropriate
constant (\eg\ X4, to provide better comparison with the source
profile), they are plotted according to the observer's perspective:
front ($\varphi^{+1}$): solid line or back ($\varphi^{-1}$): dotted
line. \label{figexp1} }
\end{figure}

\begin{figure}
\figurenum{2}

\centerline{\psfig{figure=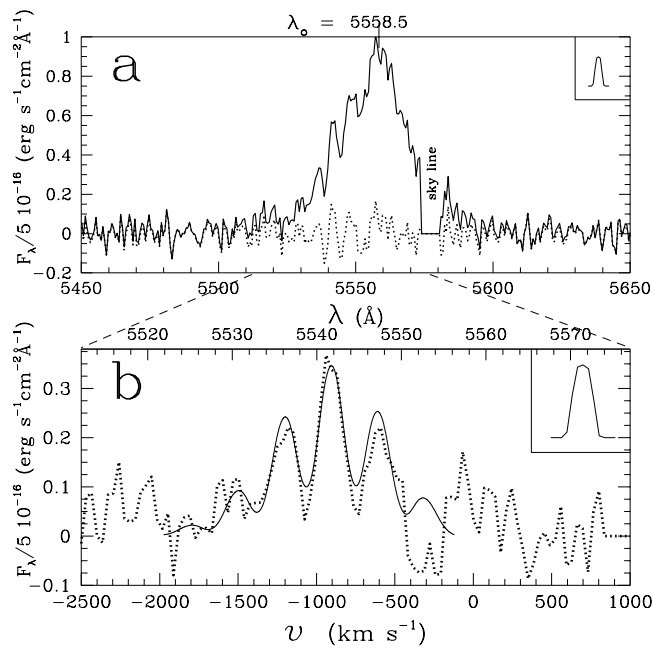,height=8.7cm,width=8.7cm}}
\figcaption{{\it Top:} The observed \lya\ profile of 1243+036 (solid
line). The dotted line represents the high frequency structures
extracted by smoothing and subtraction. {\it Bottom:} The difference
between the left half and the smoothed and folded right half of the
\lya\ profile is shown as dotted line. The lower ordinate is the
velocity shift if we adopt as line center $\lambda_0=5558.5$\AA. The
solid line represents the Fermi acceleration of the \lya\ profile of
an \hii\ region in its champagne phase. This profile has been
convolved by the instrumental profile shown in the inset
boxes. \label{figexp2} }
\end{figure}


\begin{deluxetable}{crrrrrr}
\tablenum{1}
\scriptsize
\tablecaption{Parameters characterizing profiles $\varphi^{+1}$,
$\varphi^{-1}$ of Fig.~1.} \label{tbl-1}
\tablewidth{0pt}
\tablehead{
\colhead{Model} & \colhead{\vs } & \colhead{\nsou }   & 
\colhead{\nmir \tablenotemark{a} }   & \colhead{\tsou } & 
\colhead{\tmir}  & \colhead{\mud\tablenotemark{b} } 
}

\startdata

a & 3.0 & 0.01    & 0.1   & 1.0   & 1.0 & 0  \nl
b & 3.0 & 0.2     & 0.2   & 1.0   & 1.0 & 0  \nl
c & 3.0 & 0.1     & 1.0   & 1.0   & 1.0 & 0  \nl
d & 3.0 & 1.0     & 1.0   & 1.0   & 1.0 & 0  \nl
e & 3.0 & 5.0     & 1.0   & 1.0   & 1.0 & 0  \nl
f & 3.0 & 1.0     & 0.1   & 1.0   & 1.0 & 0  \nl
g & 3.0 & 0.1     & 0.01  & 1.0   & 1.0 & 0  \nl
h & 3.0 & 0.1     & 0.001 & 1.0   & 1.0 & 0  \nl
i & 3.0 & 3.0     & 1.0   & 1.0   & 1.0 & 0  \nl
j & 3.0 & 3.0     & 1.0   & 1.0   & 1.0 & 0.2\tablenotemark{c}\nl
k & 3.0 & 0.0002  & 1.0   & 100   & 1.0 & 0  \nl
l & 3.0 & 0.28    & 0.53  & 1.0   & 1.0 & 0  \nl
\enddata
 
\tablenotetext{a}{We have set $\nprec=\nmir$ in all the above
calculations.}  \tablenotetext{b}{The dust-to-gas ratio within the two
mirrors relative to the local interstellar value (\mud=1).}
\tablenotetext{b}{The fraction of \lya\ absorbed by dust within the
two mirrors is 63\%.}

\end{deluxetable}


\begin{thebibliography}{}

\bibitem[\protect\citename{Joguet \& Binette }1998]{bj}
Joguet, B., \& Binette, L. 1998, A\&A, in preparation

\bibitem[\protect\citename{Binette \etal\ }1993]{bwzm} Binette, L.,
Wang, J. C. L., Zuo, L., \& Magris, G. 1993, AJ, 105, 797

\bibitem[\protect\citename{Bonilha \etal\ }1979]{bo79} Bonilha, J. R. M.,
Ferch, R., Salpeter, E. E., Slater, G., \& Noerdlinger, P. D. 1979, ApJ, 233, 649 


\bibitem[\protect\citename{Ferruit \etal\ }1997]{febi97} Ferruit P.,
Binette, L., Sutherland, R. S., \& P\'econtal, E. 1997, A\&A, 322, 73


\bibitem[\protect\citename{Lang }1974]{la74} Lang, K. R. 1974, in
Astrophysical formulae, Springer Verlag, Berlin



\bibitem[\protect\citename{Neufeld \& McKee }1988]{nm88} 
Neufeld, D. A., \& McKee, C. F. 1988, ApJ, 331, L87 (NM88)

\bibitem[\protect\citename{Rees}1989]{re89} 
Rees, M. J. 1989, MNRAS, 239, 1P



\bibitem[\protect\citename{Tenorio-Tagle}1979]{tt79} 
Tenorio-Tagle, G. 1979, A\&A, 71, 59



\bibitem[\protect\citename{van Ojik}1995]{vo95} van Ojik R. 1995,
PhD thesis, Leiden University (The Netherlands)

\bibitem[\protect\citename{van Ojik \etal\ }1996]{vo96} van Ojik, R.,
R\"ottgering, H. J. A., Carilli, C. L., Miley, G. K., Bremer, M. N.,
Macchetto, F. 1996, A\&A, 313, 25 (vO96)

\bibitem[\protect\citename{van Ojik \etal\ }1997]{vo97} van Ojik R., 
R\"ottgering, H. J. A., Miley, G. K., \& Hunstead, R. W. 1997, A\&A,
317, 358





\end{thebibliography}
\end{document}